\documentclass[12pt]{article}
\usepackage{epsf}

\usepackage{epsfig,graphics}
\usepackage {graphicx}
\usepackage {epsfig}
\usepackage {psfig}
\usepackage {subfigure}
\usepackage {tabularx} 
\usepackage{rotate}	
\usepackage{slashed}

\usepackage{amsmath}
\usepackage{amsfonts}
\usepackage{amssymb}
\usepackage{graphicx}

\newcommand{\bmat}{\left(\begin{array}}
\newcommand{\emat}{\end{array}\right)}

\def\yzero{\smash{\hbox{$y\kern-4pt\raise1pt\hbox{${}^\circ$}$}}}

\def\beq{\begin{equation}}
\def\eeq{\end{equation}}
\def\beqa{\begin{eqnarray}}
\def\eeqa{\end{eqnarray}}

\def\-{\hphantom{-}}

\def\s2{\frac{1}{\sqrt2}}

\def\beq{\begin{equation}}
\def\eeq{\end{equation}}
\def\beqa{\begin{eqnarray}}
\def\eeqa{\end{eqnarray}}

\def\IF{\relax{\rm I\kern-.18em F}}
\def\II{\relax{\rm I\kern-.18em I}}
\def\IP{\relax{\rm I\kern-.18em P}}
\def\IC{\relax\hbox{\kern.25em$\inbar\kern-.3em{\rm C}$}}
\def\IR{\relax{\rm I\kern-.18em R}}

\def\Dsl{\,\raise.15ex\hbox{/}\mkern-13.5mu D} 
\def\IZ{Z\kern-.4em  Z}

\catcode`\@=11   
\newdimen\@rotdimen
\newbox\@rotbox  

\def\@vspec#1{\special{ps:#1}}
\def\@rotstart#1{\@vspec{gsave currentpoint currentpoint translate
   #1 neg exch neg exch translate}}
\def\@rotfinish{\@vspec{currentpoint grestore moveto}}
\def\@rotr#1{\@rotdimen=\ht#1\advance\@rotdimen by\dp#1%
   \hbox to\@rotdimen{\hskip\ht#1\vbox to\wd#1{\@rotstart{90 rotate}%
   \box#1\vss}\hss}\@rotfinish}
\def\@rotl#1{\@rotdimen=\ht#1\advance\@rotdimen by\dp#1%
   \hbox to\@rotdimen{\vbox to\wd#1{\vskip\wd#1\@rotstart{270 rotate}%
   \box#1\vss}\hss}\@rotfinish}%
\def\@rotu#1{\@rotdimen=\ht#1\advance\@rotdimen by\dp#1%
   \hbox to\wd#1{\hskip\wd#1\vbox to\@rotdimen{\vskip\@rotdimen
   \@rotstart{-1 dup scale}\box#1\vss}\hss}\@rotfinish}%
\def\@rotf#1{\hbox to\wd#1{\hskip\wd#1\@rotstart{-1 1 scale}%
   \box#1\hss}\@rotfinish}%
\def\rotate{\@ifnextchar[{\@rotate}{\@rotate[l]}}
\def\@rotate[#1]#2{\setbox\@rotbox=\hbox{#2}\@nameuse{@rot#1}\@rotbox}

\catcode`\@=12

\topmargin
-1.5cm
\textwidth
15.5cm
\textheight
23.5cm
\oddsidemargin
0.7cm
\evensidemargin
0.7cm

\begin{document}

\makeatletter
\@addtoreset{equation}{section}
\makeatother
\renewcommand{\theequation}{\thesection.\arabic{equation}}
\pagestyle{empty}
\rightline{ IFT-UAM/CSIC-13-008}
\rightline{FTUAM-13-126}
\vspace{1.5cm}
\begin{center}


\LARGE{The Higgs Mass as a Signature of  Heavy SUSY 
} 
  \\[3mm]
  
  \vspace{2.0cm}
  
 \large{ Luis E. Ib\'a\~nez  and Irene Valenzuela\\[3mm]}
\small{
 Departamento de F\'{\i}sica Te\'orica 
and Instituto de F\'{\i}sica Te\'orica  UAM-CSIC,\\[-0.3em]
Universidad Aut\'onoma de Madrid,
Cantoblanco, 28049 Madrid, Spain 
\\[12mm]}
\small{\bf Abstract} \\[7mm]
\end{center}
\begin{center}
\begin{minipage}[h]{15.0cm}
We compute the mass of the Higgs particle in a  scheme in which SUSY 
is broken at a large scale $M_{SS}$ well above the electroweak scale $M_{EW}$. 
Below $M_{SS}$ one assumes one is  just  left with the SM with a fine-tuned Higgs potential.
Under standard unification assumptions one can compute the mass of the Higgs particle as a function of the 
SUSY breaking scale $M_{SS}$. For $M_{SS}\gtrsim  10^{10}$ GeV one obtains $m_H=126\pm 3$ GeV,
consistent with CMS and ATLAS results.
For lower values of $M_{SS}$ the values of the Higgs mass tend to those of a fine-tuned MSSM with
$m_H\lesssim130$ GeV. 
 These results support  the idea that the measured value of the Higgs mass at LHC may be
considered as indirect evidence for  the existence of  SUSY  at some  (not necessarily low) 
mass scale.

\end{minipage}
\end{center}
\newpage
\setcounter{page}{1}
\pagestyle{plain}
\renewcommand{\thefootnote}{\arabic{footnote}}
\setcounter{footnote}{0}



\section{Introduction}

The evidence \cite{:2012gu},\cite{:2012gk}
obtained by the CMS and ATLAS experiments at CERN of a scalar particle 
with the properties of a Standard Model (SM) Higgs particle with mass 
$m_H\simeq 126$ GeV is a crucial piece of information to unravel the 
origin and characteristics of the Electroweak (EW) symmetry breaking.
This mass value is compatible with the region allowed by the  MSSM
which is $m_H\lesssim 130$ GeV. Still  getting a value of the Higgs mass of
order 125 GeV in the MSSM requires a certain amount of fine-tuning.
 On the other hand 
within the SM any value from the LEP bound up to almost 1 TeV could have been possible.
Thus one might interpret the experimental results as pointing in the direction of 
some sort of (fine-tuned) SUSY.

Building on ideas discussed in 
\cite{imrv}, in  the present paper we try to answer the following question. Imagine the SM is extended 
to the MSSM above a certain scale $M_{SS}$ not necessarily tied to the EW scale, but possibly
much higher. If that is the case, a fine-tuning of the underlying theory would be required
in order for a Higgs doublet  to remain massles. Under those circumstances, what would be the 
mass of the fine-tuned Higgs?

Although the question sounds too generic to have a sharp answer it turns out that 
under standard unification assumptions a concrete answer may be given. In particular,  
assuming the unification of Higgs mass parameters
$m_{H_u}=m_{H_d}$ at the GUT/String scale  and a minimally 
 fine-tuned Higgs below the SUSY breaking scale $M_{SS}$, then  
one obtains a definite prediction for the Higgs mass as a function of the SUSY breaking scale.
Although the experimental error from the top quark mass as well as the SUSY spectra introduce  some degree of uncertainty,
the results, exemplified in fig.\ref{plotguay}, show that for $M_{SS}\gtrsim  10^{10}$ GeV the value of the Higgs mass 
is centered around $126\pm 3$ GeV. Below that scale this mass depends more on the details of the 
SUSY breaking mass parameters  but the maximum value is bound by 130 GeV, corresponding to a standard
fine-tuned MSSM with $M_{SS}\simeq 10-100$ TeV.

This predictivity  is remarkable, since the SM by itself would allow  for a large range of consistent values 
with e.g. much higher values for the Higgs mass of order e.g.  150-300 GeV or higher.
The fact that experimentally $m_H\simeq 126$ GeV then renders strong support to 
the idea of SUSY being realized at some, possibly large, mass scale. Even if SUSY particles 
are not found at LHC energies the particular value of the Higgs mass points to an underlying 
SUSY at some higher scale. This is of course due to the fact that SUSY, even spontaneously broken
at an arbitrarily high energy scale, 
relates de quartic Higgs selfcoupling to the EW gauge couplings.

\section{Traces  from high energy SUSY and a minimally fine-tuned Higgs}

There is at present no experimental evidence at LHC for the existence of SUSY particles. This,  combined with
earlier experimental limits,  severely  constraint the idea of low energy SUSY and indicates the necessity
of some degree of fine-tuning of parameters of the order of at least 1-0.1 percent \cite{fine-tuninghiggs,aci2}. If no evidence of SUSY particles 
is found at the 14 TeV LHC the general idea of low-energy SUSY as a solution to  the hierarchy/fine-tuning problem
will be strongly questionable. On the other hand, 
as recently emphasized in \cite{imrv}, 
even  if SUSY is not present at the EW scale to solve the hierarchy problem, there are at least three 
reasons which suggest  that supersymmetry could be present at some scale $M_{SS}$ above the EW scale and below
the unification/string scale. The first is the fact that SUSY is a substantial ingredient of string theory which is, as of today,
the only serious contender for an ultraviolet completion of the SM.  The second reason is that SUSY guarantees the 
absence of scalar tachyons which are generic in non-SUSY string vacua. Thirdly, and independently from any string
theory consideration, a detailed  study of the non-SUSY SM Higgs potential consistent with the measured
Higgs mass indicates that there is an instability at scales above $\simeq 10^{10}$ GeV \cite{previos,recien}. Although in principle one
can live in a metastable vacuum, supersymmetry would  stabilize the vacuum in a natural way if
present at an energy scale $\lesssim 10^{10}$ GeV.

Let us then consider a situation in which SUSY is broken at some high scale $M_{SS}$ with $M_{EW}\ll M_{SS}\ll M_{C}$,
where $M_C$ is the unification/string  scale. For previous work on a  fine-tuned Higgs in a setting with broken SUSY at a  high
scale see e.g.\cite{splitsusy,Hall:2009nd,otheranthropic,Hebecker:2012qp,imrv,Arvanitaki:2012ps}.
With generic SUSY breaking soft terms one is just left at low-energies
with the SM spectrum. In addition the scalar potential should be fine-tuned so that a Higgs doublet remains light
so as to trigger EW symmetry breaking. 
One would say that no trace would be left from the underlying 
supersymmetry. However this is not the case \cite{Hall:2009nd}. Since dimension four operators are not affected by
spontaneous SUSY breaking, the value $\lambda(M_{SS})$ of the Higgs self-coupling at the $M_{SS}$ scale  will be given 
in the MSSM by the  (tree level) boundary condition
\beq
\lambda_{SUSY}(M_{SS}) \ =\ \frac {1}{4}
(g_2^2 \ +\  g_1^2) \ cos^22\beta
\label{lambdasusy}
\eeq
which is inherited from the D-term scalar potential of the MSSM.
Here $g_{1,2}$ are the EW gauge couplings and $\beta$ is the mixing angle 
which defines the linear combination of the two $SU(2)$ doublets $H_u,H_d$ 
of the MSSM which remains massless after SUSY breaking, i.e.,
\beq
H_{SM} \ =\ sin\beta H_u \ +\  cos\beta H_d^* \ .
\eeq
Thanks to this boundary condition, for any given value of  tan$\beta$ one can compute 
the Higgs mass as a function of the SUSY breaking scale $M_{SS}$. 

Schematically the idea is to run  in energies the values of $g_1,g_2$ up to the given $M_{SS}$ scale. For any value of tan$\beta$ 
one then computes $\lambda(M_{SS})$ from eq.(\ref{lambdasusy}). Starting with this value we then run down in energies
and obtain the value for the Higgs mass  from $m_H^2(Q)= 2v^2\lambda(Q)$. Threshold corrections at both the EW and SUSY scales 
have to be included. This type of computation  for different values of tan$\beta$ was done e.g  in 
ref.\cite{Cabrera:2011bi}, \cite{Giudice:2011cg}, \cite{Arbey:2011ab}, \cite{EliasMiro:2011aa}. 
    We show results for a similar computation
in fig.\ref{plotguay} (grey bands). The Higgs mass may have any value in a broad band below a maximum around 140 GeV. 
One may easily understand the general structure of these curves. The mass is higher for higher tan$\beta$ since the
tree level contribution to the Higgs mass through eq.(\ref{lambdasusy}) is higher. On the other hand the Higgs mass slowly grows 
with larger $M_{SS}$ as expected.

 What we want to emphasize here is that the natural  assumption of Higgs soft mass unification 
 at the unification  scale $M_C$, i.e.
 \beq
 m_{H_u}(M_C) \ =\ m_{H_d}(M_C)
 \label{unifmasses}
 \eeq
 leads to a much more restricted situation with trajectories in the 
 $m_{higgs}-M_{SS}$ plane rather than a wide band. Note that this equality is quite
 generic in most SUSY, unification or string models. In particular it appears in gravity mediation
 as well as in almost all SUSY breaking schemes, including those arising from
 compactified string theory, see e.g.\cite{BOOK}.

Indeed, to see this let us recall what is the 
general form for  Higgs masses in the MSSM at the scale $M_{SS}$, 
 \beq
\left(
\begin{array}{cc}
 {{H_u}} \!\!\! & ,\  {{ H}_d^*}\\
\end{array}
\right)
\left(
\begin{array}{cc}
 { m_{H_u}^2} &   m_{3}^2\\   
  { m_3^2} & {m_{H_d}^2 }\\
\end{array}
\right)
\left(
\begin{array}{c}
  {{ H_u^*}} \\
  {{ H}_d}\\
\end{array}
\right) \,.
\label{matrizmasas}
\eeq
where we will take $m_3^2$ real. 
If all these mass terms were zero we would get two Higgs doublets
in the massless spectrum. However this would require  extra unnecessary fine-tuning. 
The {\it minimal Higgs fine-tuning} would only require a single Higgs doublet 
to remain at low-energies \footnote{This is a particular realization of the {\it Extended Survival Hypothesis} of ref.\cite{ai}(see also \cite{dg}).}
.  This is achieved for a single   fine-tuning 
$m_3^4=m_{H_u}^2m_{H_d}^2$. The massless eigenvector is then
\beq 
H_{SM} \ =\ sin\beta \ H_u \ +\ cos\beta \ H_d^*
\eeq
with 
\beq
tan\beta \ =\ \frac {|m_{H_d}|}{|m_{H_u}|} \ .
\label{tanbetacal}
\eeq
If the origin of this fine-tuning is understood in terms of the  fundamental SUSY breaking parameters {\it scanning} 
in a {\it landscape} of possibilities, 
 the diagonal Higgs masses are supposed to {\it scan}  in a way consistent with the boundary 
condition  (\ref{unifmasses}).
One can then compute the value of tan$\beta(M_{SS})$ by running the ratio in (\ref{tanbetacal})  from the
unification scale $M_C$ down to the SUSY breaking scale $M_{SS}$. One computes  the value of the
Higgs self-coupling $\lambda(M_{SS})$ from eq.(\ref{lambdasusy}) and then runs  down in energies to compute the
Higgs mass for any given value of $M_{SS}$.
In a general MSSM model  we can compute this  in terms of the underlying structure of soft terms at $M_{SS}$.  
In particular one expects generic SUSY-breaking soft terms of order $M_{SS}$. For definiteness we  will 
assume here a universal structure of soft terms with the standard parameters $m$ (3-d generation scalars masses), $M$ (gaugino masses),  $A$
 (3-d generation trilinear parameter)
and $\mu$ (mu-term). As we will see, the results will have very little dependence on this universality assumption which simplifies  substantially 
the computations.  This universality assumption is also consistent with the (weaker) assumption of Higgs mass unification, eq.(\ref{unifmasses}).

Let us remark that in this approach  the only relevant condition is $m_{H_u}=m_{H_d}$ at the unification scale $M_C$. There is no  need
for a {\it  shift symmetry}  which imposses $m_3^4=m_{H_u}^ 2m_{H_d}^ 2$ at the unification scale as in ref.\cite{Hebecker:2012qp},  since then the fine-tuning would be 
destroyed by the running from $M_C$ to $M_{SS}$. The idea is that enviromental selection should ensure that at the scale $M_{SS}$ (not $M_C$) the fine-tuning 
condition $m_3^4=m_{H_u}^ 2m_{H_d}^ 2$ is impossed with high accuracy.

\section{The Higgs mass from minimal fine-tuning}

We now turn to a description of the different steps required to compute the Higgs mass as a function of the 
SUSY-breaking scale $M_{SS}$.

\subsection{Computing the couplings at $M_{SS}$}

We start by computing the electroweak couplings at the $M_{SS}$ scale. We take the central values 
for  the masses (in GeV) and couplings at the weak scale
\beq
M_Z=91.1876\ ,\ 
M_W=80.385  \ ,\ 
m_t=173.1\ 
\label{input1}
\eeq
\beq
\sin^2\theta_W(M_Z)=0.23126  \ ,\ 
\alpha_{em}^{-1}(M_Z)=127.937 \ ,\ 
\alpha_3(M_Z)=0.1184 \ .
\label{input2}
\eeq
We will allow to vary  the top mass with an error  $m_t=173.1\pm 0.7$ GeV  obtained from the average  from 
Tevatron  \cite{Aaltonen:2012ra} and CMS and ATLAS results as in ref.\cite{espinosa}.
 We will  neglect the error from $\alpha_3$ which is much smaller than that from the 
top quark mass.  To extract the value of the top Yukawa coupling $h_t(m_t)$ we take into account the relationship 
between the pole top-quark mass $m_t$ and the corresponding Yukawa coupling in the  $\overline{MS}$ scheme
\cite{Chetyrkin:1999qi}
\begin{equation}
h_t(m_t)=\frac{m_t}{v}(1+\delta_t)
\label{gtm}
\end{equation}
where the dominant one-loop QCD corrections may be estimated (\cite{Chetyrkin:1999qi}, \cite{Giudice:2011cg,EliasMiro:2011aa})
\begin{equation}
\delta_{t}^{QCD}(m_t)=-\frac{4}{3 \pi}\alpha_3(m_t)-0.93 \alpha_3^2(m_t)-2.59\alpha_3^3(m_t)\approx -0.0605 \ .
\end{equation}
One then obtains $h_t(m_t)=0.934$.
We run now the couplings $g_1$,$g_2$ and $h_t$ up to the given scale $M_{SS}$. We do this by solving the 
RGE at two loops for the SM couplings.  Those equations are shown for completeness in
appendix A.

\subsection{Computing 	tan$\beta$ and $\lambda(M_{SS})$}

With $g_{1,2}(M_{SS})$ at hand we  want  now to compute  the value of $\lambda(M_{SS})$ from eq.(\ref{lambdasusy}).
To do that we need to compute tan$\beta(M_{SS})$ from eq.(\ref{tanbetacal}), which in turn requires the computation of the running of the 
masses $m_{H_u}$,$m_{H_d}$ from the unification scale at which $m_{H_u}=m_{H_d}$ down to $M_{SS}$. 

The value of the unification scale $M_C$ is usually obtained from the unification of gauge
coupling constants. In our case, with two regions respectively with the SM (below $M_{SS}$) and the MSSM (in between $M_{SS}$ and
$M_C$) the value of $M_C$ is not uniquely  determined. In fact it is well known that 
precise unification is only obtained for $M_{SS}\simeq 1$ TeV, as in standard MSSM phenomenology \cite{unification}.
However, approximate unification around a scale $M_C\simeq 10^{14}-10^{15}$ GeV is anyway obtained 
for much higher values of $M_{SS}$, even in the limiting case with $M_{SS}\simeq M_C$ in which case 
SUSY is broken at the unification scale, so a simple approach would be to take $M_C\simeq 10^{15}$ 
GeV to compute the runing of tan$\beta$.  We find more interesting instead to achieve consistent gauge
coupling unification from appropriate threshold corrections. 
In particular, in a large class of string compactifications like F-theory $SU(5)$ GUT's there
are small threshold corrections respecting the  boundary condition at the GUT scale
\cite{Ibanez:1998rf,Blumenhagen:2008aw,imrv}
\begin{equation}
\frac{1}{\alpha_1(M_C)}=\frac{1}{\alpha_2(M_C)}+\frac{2}{3\alpha_3(M_C)} \ .
\label{cond_unif}
\end{equation}	
This boundary condition is consistent (but more general) than the usual GUT boundary conditions
$\alpha_3=\alpha_2=5/3\alpha_1$. It arises for example from F-theory $SU(5)$ GUT's \cite{ftheoryreviews} once
fluxes along the hypercharge direction are added to break the $SU(5)$ symmetry down to the SM \cite{Blumenhagen:2008aw,DW}.
Using the RGE for gauge couplings in both SM and MSSM regions (at two loops for the gauge couplings and one loop for the top Yukawa) one finds that unification 
of couplings is neatly obtained at a scale $M_C$ related with $M_{SS}$ by the approximate 
relationship
\begin{equation}
\log{M_C} = -0.23 \log{M_{SS}} + 16.77 \ .
\label{unif}
\end{equation}
As one varies $M_{SS}$ in the range $1$ TeV-$M_C$ one obtains $M_C\simeq 10^{16}-10^{14}$ GeV.
 This relation changes very  little  compared  to the one obtained just using the RGE at one loop  in ref.\cite{imrv}.
To compute tan$\beta(M_{SS})$ we will use as unification scale the $M_C$ obtained from eq.(\ref{unif}) consistent
with gauge coupling unification. It is important to remark though that this has very little impact in the numerical results obtained,
there is no detail dependence on the value of $M_C$ as long as it remains in the expected $10^{14}-10^{17}$ GeV region.

To compute tan$\beta$ at $M_{SS}$ one solves the  RGE for the Higgs mass parameters $m_{H_u},m_{H_d}$.
At this point one needs to make some assumptions about the structure of the SUSY-breaking soft terms of 
the underlying MSSM theory. We will thus assume a standard universal SUSY breaking structure parametrized by 
universal scalar masses $m$, gaugino masses $M$ and trilinear parameter $A$. The results are independent from the 
value of the B parameter which is fixed by the fine-tuning condition at $M_{SS}$.  
Given these uncertainties it is enough to use the one-loop RGE for the soft parameters,  which were analytically solved in ref.\cite{ilm}.
As described in  \cite{imrv} one has
tan$\beta(M_{SS})=|m_{H_d}(M_{SS})|/|m_{H_u}(M_{SS})|$ with
\beq
m_{H_d}^2(t) \ =\ m^2 \ +\ \mu^2 q^2(t)\ +\ M^2 g(t)
\nonumber
\eeq
\beq
m_{H_u}^2(t)\ =\ m^2h(t)-k(t)A^2\ +\ \mu^2q^2(t)\ +\ 
M^2e(t)\ +\ AMf(t)
\label{ilm2}
\eeq
where $m,M,A,\mu $ are the standard universal CMSSM  parameters  at the unification scale $M_C$, $t=2log(M_C/M_{SS})$  
and $q,g,h,k,e,f$ are known functions of the top Yukawa coupling $h_t$ and the three 
SM gauge coupling constants. Except for regions with large tan$\beta$, appearing only for low $M_{SS}$, 
one can safely neglect the bottom and tau Yukawa couplings, $h_b=h_\tau=0$.  For completeness these functions are provided in Appendix B.
The value taken for $h_t$ to perform the running of soft terms is a bit subtle since at  $M_{SS}$ one has to match the $h_t^{SM}$ value obtained 
from the SM running up to $M_{SS}$ with the SUSY value $h_t^{SUSY}$ which are related by
\beq
h_t^{SM}\ =\ h_t^{SUSY} sin\beta \ 
\label{match}.
\eeq
Since the value of $h_t^{SUSY}$ depends on $\beta$  through eq.(\ref{match}), the computation of tan$\beta$ is done in a self-consistent way:
a value is given to sin$\beta (M_{SS})$,  $h_t^{SUSY}$  is run up in energies and one has a tentative $h_t(M_C)$. 
One then runs $m_{H_u}/m_{H_d}$ down  in energies and computes tan$\beta$ at $M_{SS}$. When both values for $\beta$
at $M_{SS}$  agree the computation of tan$\beta$ is 
consistent. 

Once computed the value of tan$\beta$ as described above, one then obtains $\lambda(M_{SS})$ from eq.(\ref{lambdasusy}).
In addition there are threshold corrections at $M_{SS}$ induced by loop diagrams involving the SUSY particles.
The leading one-loop correction is given by
\beq
\delta\lambda (M_{SS})\ =\ 
\frac{1}{(4 \pi )^2}3 h_{t}^4 \left(2 X_t-\frac{X_t^2}{6}\right)
\label{label}
\eeq
where $h_t$ is the SUSY top Yukawa coupling at $M_{SS}$ and the stop mixing parameter $X_t$ is given by 
\beq
X_t\ =\  \frac{ (A_t-\mu\cot\beta)^2}{m_Q m_U} \ .
\eeq 
with $m_Q(m_U)$ the left(right)-handed stop mass.
This term comes from finite corrections involving one-loop exchange of  top squarks. There are further correction terms which are numerically 
negligible compared to this at least for not too low $M_{SS}$, in which case the SUSY spectrum becomes more spread and further 
threshold corrections become relevant, see e.g. \cite{Giudice:2011cg}.
We have computed this parameter $X_t$ using the one loop RGE for the soft parameters that are provided in Appendix B and the value of tan$\beta$ obtained above.

\begin{figure}[t]
\begin{center}
\includegraphics[width=14cm]{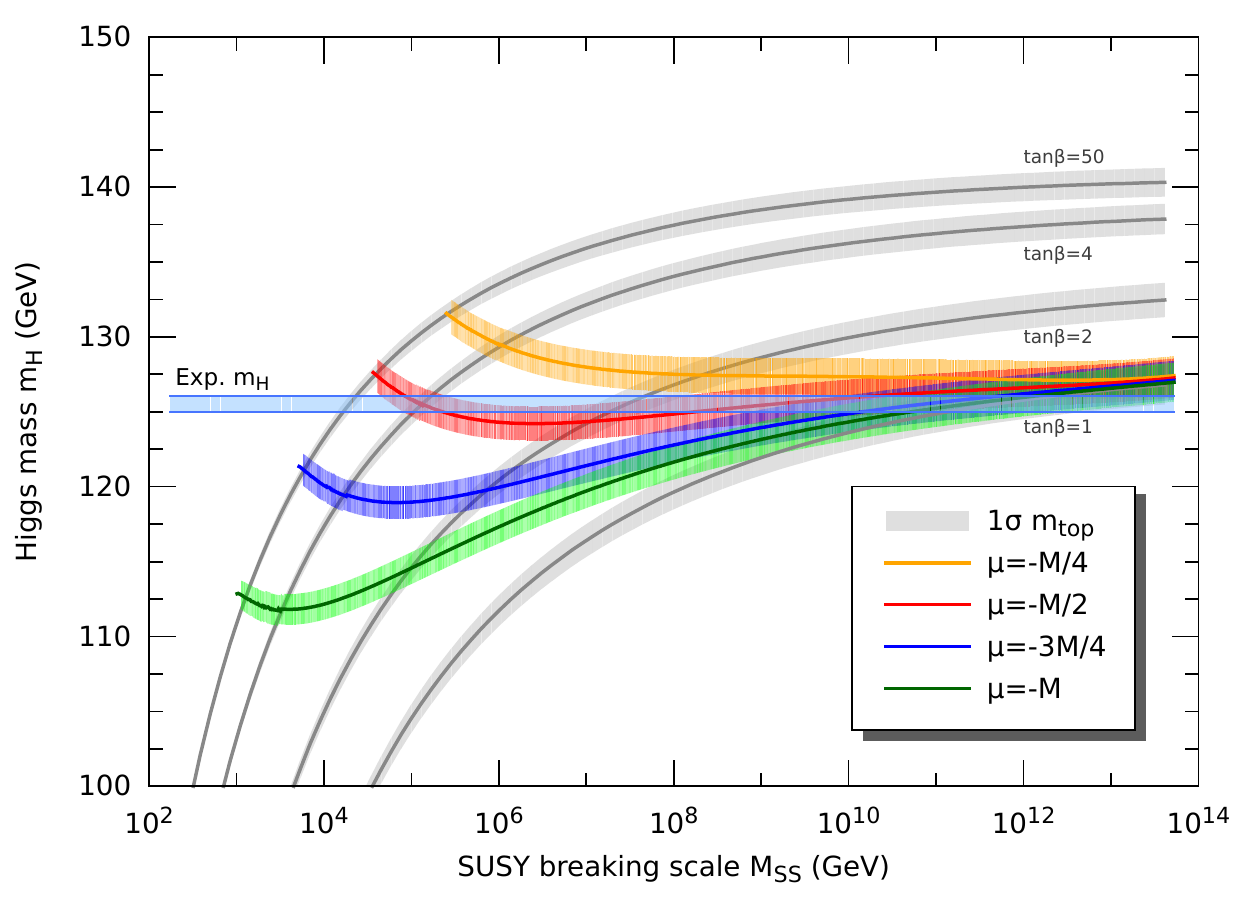}
\end{center}
\caption{\label{gcu}  Higgs mass versus SUSY breaking scale $M_{SS}$. The grey bands correspond to the Higgs mass for different values
of tan$\beta$, for $X_t=0$, without impossing unification of Higgs soft parameters. The other colored bands correspond to impossing tan$\beta$ values 
consistent with unification of soft terms, $m_{H_u}=m_{H_d}$. Results are shown for a choice of universal soft terms $M=\sqrt{2}m$, $A=-3/2M$ and four values for
the $\mu$-term. The stop mixing parameter $X_t$ is computed from the given soft parameters.
 The width of the bands correspond to the error from the top quark mass which is taken 
to be  $m_t=173.1\pm 0.7$.  The horizontal band corresponds to the ATLAS and CMS average Higgs mass result.
}
\label{plotguay}
\end{figure}
%

\subsection{Computing the Higgs mass} 

Starting  from $(\lambda +\delta \lambda)(M_{SS})$ 
 one runs back  the self-coupling 
down to the EW scale (using the SM RGE at two loops) and computes the Higgs mass at a scale $Q$  (taken as $Q=m_t$) through
\beq
m_H^2\ =\ 2 v^2(\lambda(Q)+\delta^{EW}\lambda(Q)) \ ,
\eeq
where $v=174.1$ GeV is the Higgs vev and $\delta^{EW}\lambda(Q)$ are additional EW scale threshold corrections. 
At one-loop these corrections are given by \cite{Sirlin:1985ux}
\begin{equation}
\delta^{EW} \lambda=-\frac{\lambda G_F M_Z^2}{8 \pi^2 \sqrt{2}}(\xi F_1+F_0+F_3/\xi)\approx 0.011 \lambda
\label{label}
\end{equation}
where $\xi=m_H^2/M_Z^2$ and the functions $F_1,F_0$ y $F_3$ depend only on EW parameters and are shown in appendix C  for completeness.  
This completes the computation procedure for the Higgs mass as a function of $M_{SS}$.


\begin{figure}[t]
\begin{center}
\includegraphics[width=13cm]{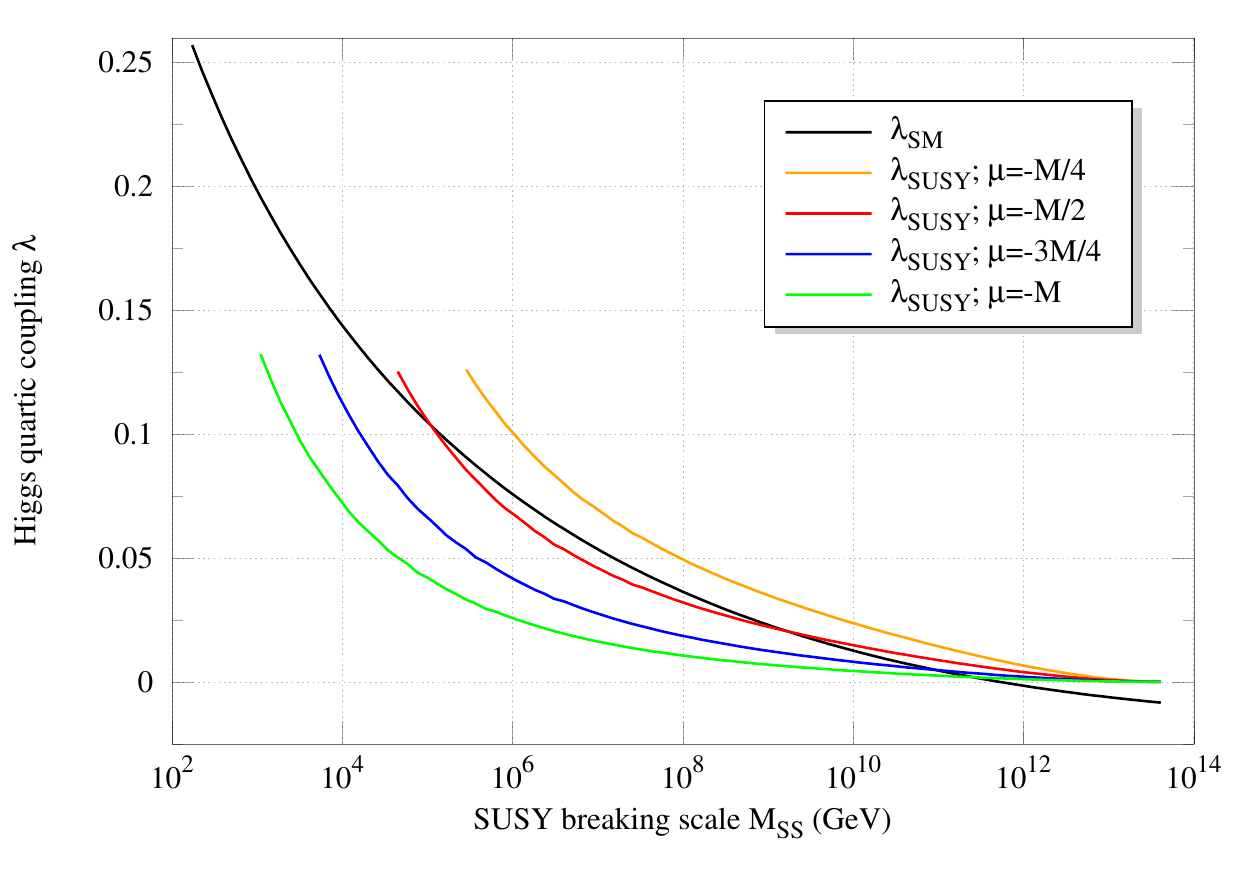}
\end{center}
\caption{\label{gcu}  The black line shows the value of the SM self-coupling 
$\lambda$ as a function of $M_{SS}$, using as input the LHC Higgs data. The remaining  curves show  values of $\lambda_{SUSY}$ consistent 
with $m_{H_u}(M_C)=m_{H_d}(M_C)$  for different values of $\mu$. When these $\lambda_{SUSY}$ lines cross the $\lambda$ curve the 
SUSY model is consistent with LHC Higgs data. 
}
\label{plotmass2}
\end{figure}
%

Figure \ref{plotguay}  plots the value of  $m_H^2$ as a function of $M_{SS}$. For definiteness we plot the results for 
universal soft terms with $M=\sqrt{2}m$, $A=-3/2M$. This choice of values is motivated by modulus dominance
SUSY breaking in string scenarios, see e.g.   \cite{Aparicio:2008wh},\cite{BOOK}.
However, as  we will explain later, other different choices for soft parameters $m,M,A$ lead to analogous results.
The grey bands correspond to the computation of the mass for 
 tan$\beta=1,2,4,50$ and $X_t=0$. The results are similar to those obtained in 
 ref.\cite{Cabrera:2011bi}, \cite{Giudice:2011cg}, \cite{Arbey:2011ab}, \cite{EliasMiro:2011aa}.
 The other colored bands correspond to the 
 Higgs mass values obtained under the assumption of Higgs parameter unification as in eq.(\ref{unifmasses}). 
 Results are displayed for  a mu-term $\mu=-M/4,-M/2,-3/4M,-M$ with the value for $X_t$ computed from the
obtained running soft terms
 \footnote{The results are very weakly dependent on the sign of $\mu$ through 
 the $X_t$ appearing in the threshold corrections.}. The width of the grey and colored bands corresponds to the 
  error from  the top quark mass. Finally the horizontal band corresponds to
 the average CMS and ATLAS results for the Higgs mass (we take $m_H=125.5\pm 0.54$, see
\cite{EliasMiro:2011aa}).

The figure shows that above a scale $\simeq 10^{10}$ GeV the value of the Higgs mass is contained in the
range
\beq
m_H \ =\ 126\ \pm\ 3 \ GeV \ .
\eeq
This is remarkably close to the measured value at LHC and supports the idea that SUSY {\it and} 
unification  underly the observed Higgs mass. This result is quite independent of the details of the soft terms.
Below $10^{9}$ GeV  the Higgs mass becomes more model dependent. In particular the Higgs mass is reduced
as   $|\mu|$ increases. This is easy to understand from eq.(\ref{ilm2}) since for larger $\mu$ the ratio $m_{H_u}/m_{H_d}$
approaches  one, yielding tan$\beta\simeq 1$.  One still gets a Higgs mass consistent with LHC results for not too large
$|\mu|$. As one approaches $M_{SS}\simeq 10-100$ TeV one reaches the region of standard fine-tuned MSSM with a
Higgs mass which may be as large as 130 GeV. As we approach that region our treatment becomes incomplete since 
some neglected SUSY threshold corrections beyond those in (\ref{label}) become important, and  the SUSY spectrum spreads out. However, that 
region corresponds to the well understood situation of the MSSM with a heavy SUSY spectrum with masses 
in the 10-100 TeV region.

Let us finally note that, within uncertainties, the  figure also  favours values for the SUSY breaking scale 
$M_{SS}\lesssim 10^{13}$ GeV.

One may also interpret graphically the above results in terms of the unification of the SM Higgs self-coupling $\lambda_{SM}$ 
and the SUSY predicted self-coupling $\lambda_{SUSY}=(g_1^2+g_2^2)cos^22\beta/4$.  This is depicted in fig.\ref{plotmass2},
in which we have not included the uncertainty from the $m_t$ error to avoid clutter. Note that the dependence of $\lambda_{SUSY}$ 
on $M_{SS}$ is qualitatively similar to the running of $\lambda_{SM}$. This may be understood as follows.
In the definition of $\lambda_{SUSY}$,  $(g_1^2+g_2^2)$ runs very little and remains practically constant. On the other hand 
one has $cos^22\beta=(m_{H_u}^2-m_{H_d}^2)^2/(m_{H_u}^2+m_{H_d}^2)^2$. The difference on the numerator goes like
$h_t^4$, which is also the order of the leading correction to the $\lambda_{SM}$ coupling.

\begin{figure}[t]
\begin{center}
\includegraphics[width=13cm]{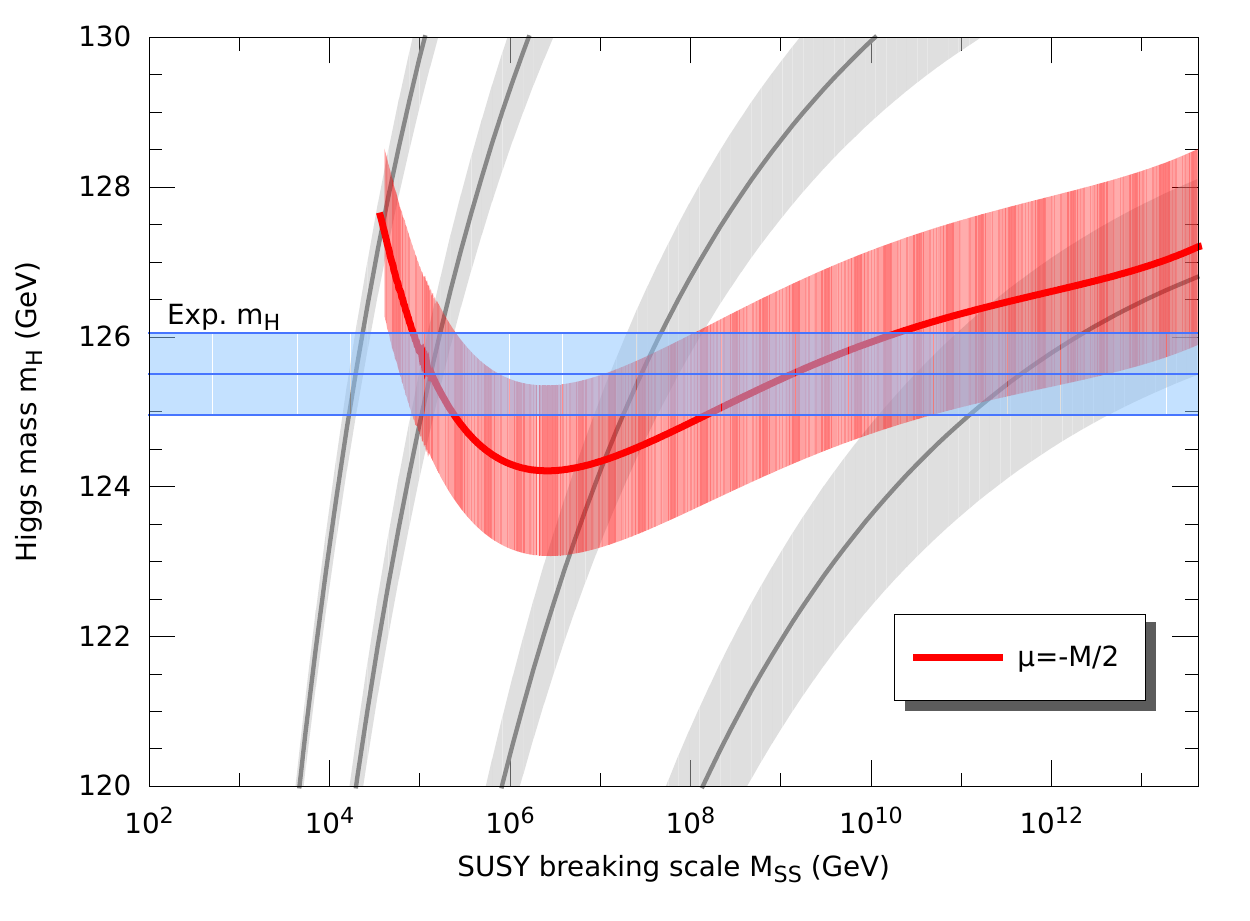}
\end{center}
\caption{\label{gcu}  Higgs mass versus SUSY breaking scale $M_{SS}$ for $\mu=-M/2$ (red band). 
Its width reflects the uncertainty on $m_t=173.1\pm 0.7$. The grey bands, as in fig.\ref{plotguay} 
show the Higgs mass for several values of tan$\beta=1,2,4,50$ and are displayed to guide the eye.
}
\label{plotmas2}
\end{figure}
%

\section{Model dependence}

In this section we discuss  different model dependent possibilities which arise depending on the structure of
the underlying soft terms. With sufficiently precise information about the top quark and Higgs masses one may obtain
interesting  constraints on the possible structure of the SUSY-breaking terms. 

\begin{figure}[t]
\begin{center}
\includegraphics[width=0.492\textwidth]{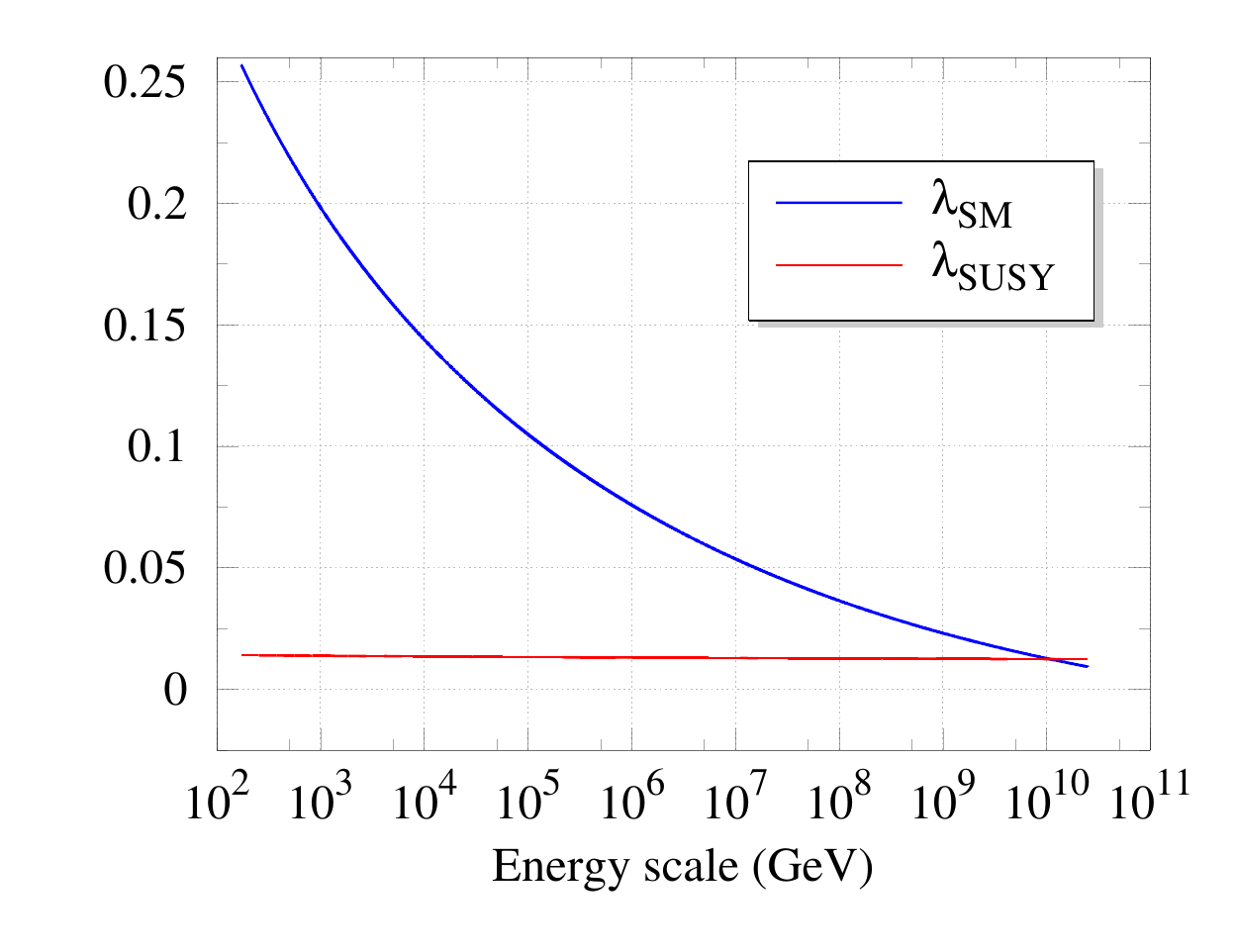} \
\includegraphics[width=0.492\textwidth]{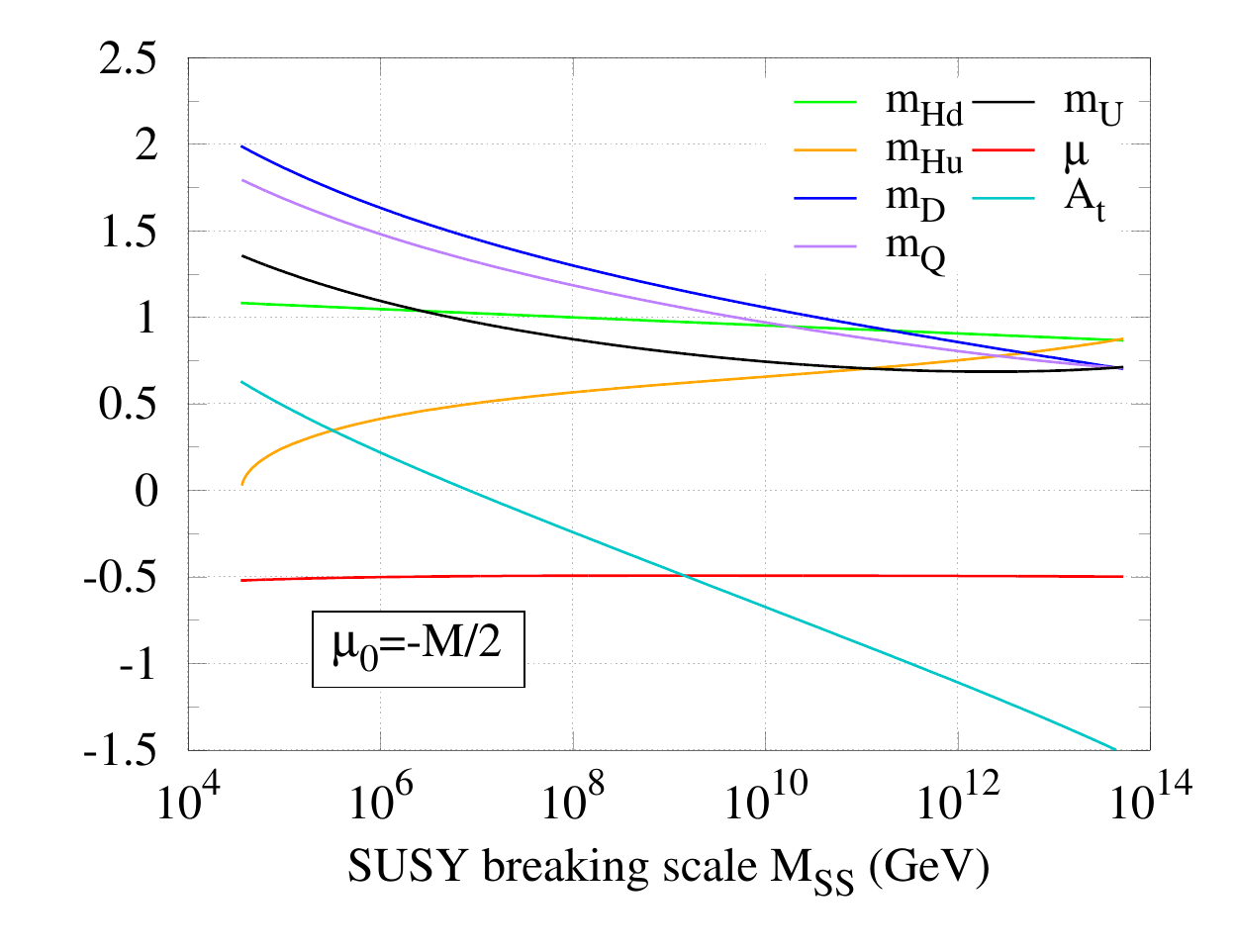}
\end{center}
\caption{\label{gcu}  Left: Evolution of the SM Higgs selfcoupling $\lambda(t)$ and the  combination 
$\lambda_{SUSY}=(g_1^2(t)+g_2^2(t))/4\times cos^2(2\beta)(M_{SS})$  in the model with $\mu=-M/2$ and an  intermediate scale $M_{SS}\approx 3\cdot10^{10}$GeV.
They unify at $M_{SS}$ where SUSY starts to hold.
Right: Values of the 3-d generation squark  soft masses $m_{Q,U,D}$ as well the Higgs
mass parameters $m_{H_u},m_{H_d},\mu$ and trilinear $A_t$  at the scale $M_{SS}$ obtained from the
running below the   unification scale $M_C$.
}
\label{plotmass3}
\end{figure}

Let us concentrate first in the case with universal soft terms and $\mu=-M/2$ but still keeping the
relationships $M=\sqrt{2}m$, $A=-3/2M$.  As we said these values are interesting since, as discussed in ref.\cite{imrv},   they 
may be understood as arising from a Giudice-Masiero mechanism in a modulus dominance SUSY breaking scheme.
The dependence of the Higgs mass as a function of $M_{SS}$ in this particular case 
is shown in fig.\ref{plotguay}  with the red band, a zoom is provided in fig.\ref{plotmas2}.
Given the uncertainties, in this particular case ($\mu=-M/2$) essentially any value for $M_{SS}$ in the 
$10^4-10^{14}$ GeV region is consistent with the observed Higgs mass, although regions around $10^4-10^5$ and 
$10^8-10^{10}$ GeV are slightly favoured. 
 This second possibility with $M_{SS}\simeq 10^{10}$ GeV was explored in \cite{imrv} (see also\cite{Hebecker:2012qp})
 in which it was argued
that such intermediate  SUSY breaking may be interesting for two additional reasons
\footnote{An additional interesting property is that for $M_{SS}\gtrsim 10^ {10}$ GeV such models do not require the implementation of doublet-triplet splitting nor 
R-parity preservation. No Polony problem is present either \cite{imrv}.}.
On one hand this scale naturally appears in string compactifications in which SUSY breaking is induced by closed string fluxes. 
Indeed in such a case one has \cite{imrv}
\beq
 M_{SS}\ \simeq \ \sqrt{2g_s/\alpha_G}(M_C^2/M_p) \ ,
 \label{mssflux}
 \eeq
 where $g_s$ is the string coupling, $\alpha_G$ is the unified fine structure
constant and $M_p$ is the Planck mass.  For $g_s\simeq 1$ and $M_C\simeq 10^{14}$ GeV one indeed gets $M_{SS}\simeq 10^{10}$ GeV.
The second reason is that in those constructions an axion with a scale $F_a\simeq M_C/(4\pi)^2\simeq 10^{12}$ GeV appears,
which is consistent with axions providing for the dark matter in the universe. In this case, using eqs.(\ref{unif}),(\ref{mssflux}) one
obtains $M_{SS}=2.49\times 10^{10}$ GeV and $M_C=2.43\times 10^{14}$ GeV.
Values this low for the unification scale can still be compatible with
proton decay constraints \cite{imrv}.
Computing the Higgs mass following the procedure described in the previous section one 
obtains in this case 
\beq
m_H \ = \ 126.1 \pm 1.2  \ \ GeV \ 
\eeq
where the error includes only that coming from the top mass uncertainty. 
This is clearly consistent with the findings at ATLAS and CMS.
In this scheme with an intermediate scale
$M_{SS}$ the Higgs self-coupling unifies with its SUSY extension as depicted in fig.\ref{plotmass3} (left) . The soft masses evolve logarithmically 
from $M_C$ down to $M_{SS}$ as depicted in fig.\ref{plotmass3} (right). The value of tan$\beta$ increases as the value of $m_{H_u}^2$ decreases
and $m_{H_d}^2$ remains almost constant, so that tan$\beta$ increases as $M_{SS}$ decreases.

%

\begin{figure}[t]
\begin{center}
\includegraphics[width=0.6\textwidth]{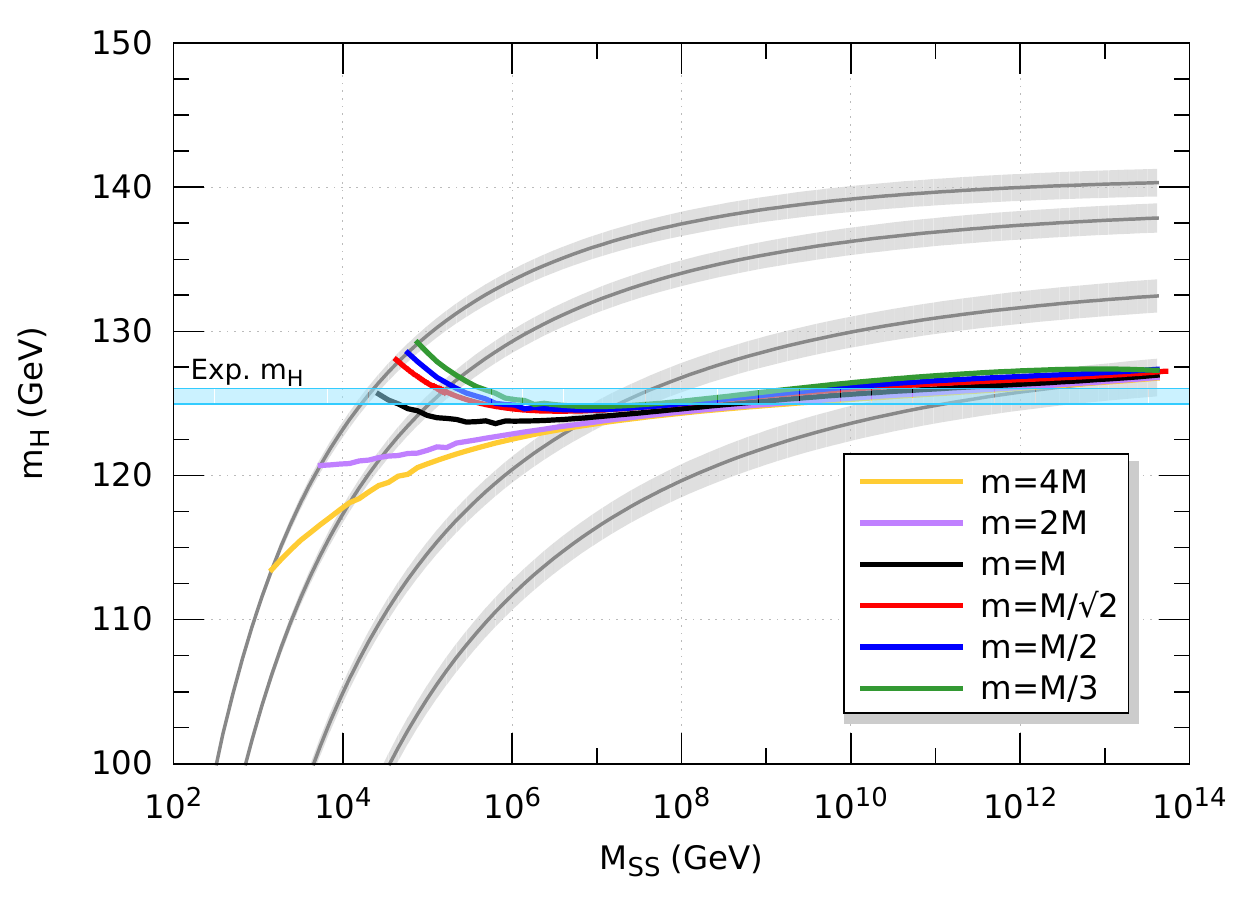} \ 
\includegraphics[width=0.6\textwidth]{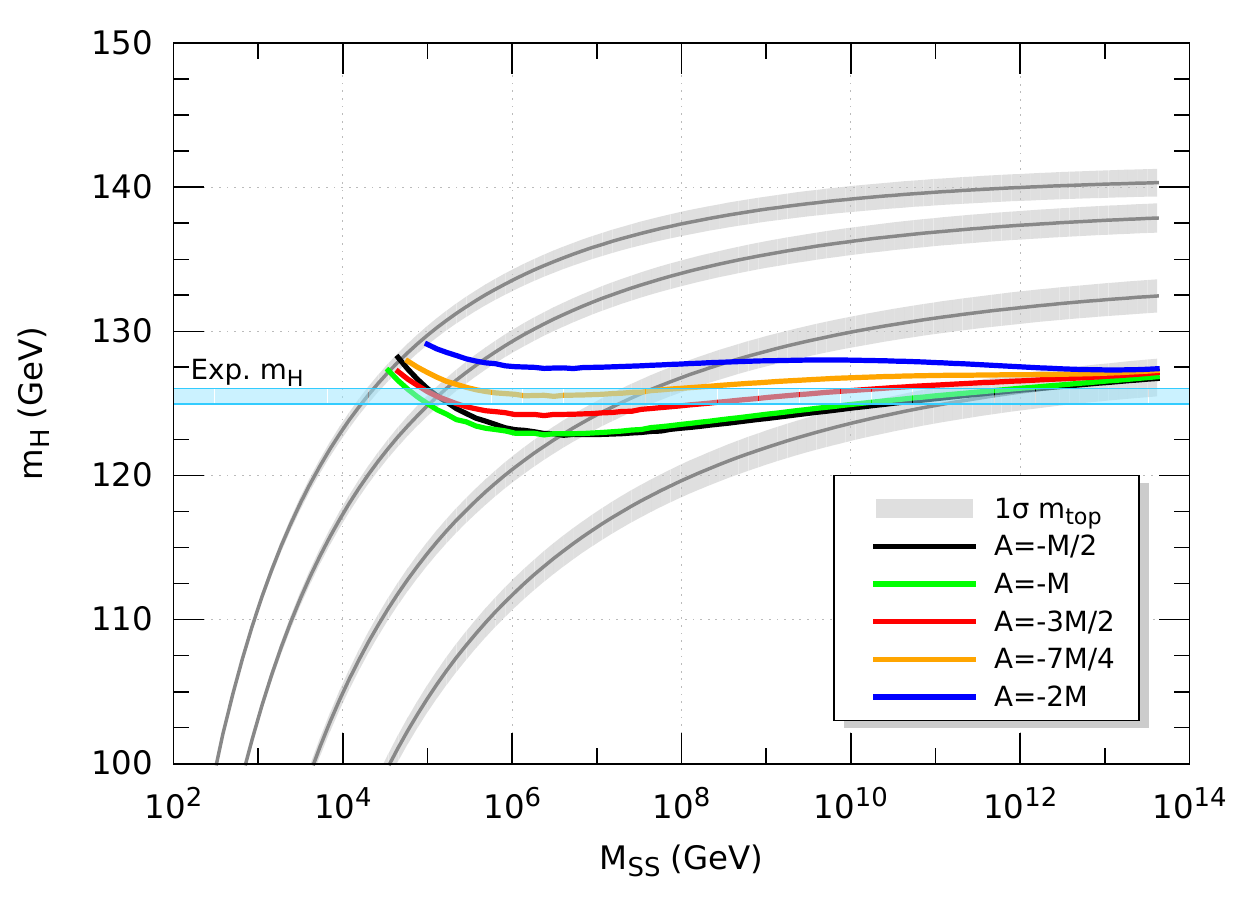} 
\end{center}
\caption{\label{gcu}  Higgs mass versus SUSY breaking scale $M_{SS}$ for $\mu=-M/2$. Up:  for various values of the scalar mass
parameter $m$ in units of the gaugino mass $M$; Down: for various values of the trilinear $A$ parameter.
}
\label{plotmass6}
\end{figure}
%

It is interesting to explore how relaxing the above mentioned relationships $M=\sqrt{2}m$, $A=-3/2M$ modify the results for the Higgs
mass.  In fig.\ref{plotmass6} (up) we show how the prediction for the Higgs mass is changed as one varies the value of $m$ away from $m=M/\sqrt{2}$.
The figure remains qualitatively the same but one observes that as $m/M$ increases the Higgs mass tends to be lighter.
Above $M_{SS}\simeq 10^7$ GeV the Higgs mass remains in the region $m_H\simeq 126\pm 3$ GeV.
The effect of varying $A$ away from $A=-3/2M$ is also shown in fig.\ref{plotmass6} (down). 
Although we have not included  the error coming from the top quark mass to avoid clutter, one concludes 
that the overall  structure remains  the same and the Higgs mass stays around $126\pm 3$ GeV for $M_{SS}\gtrsim 10^{10}$ GeV.
However now values of $M_{SS}$ in between 100 TeV and $10^{10}$ GeV are consistent with the observed Higgs mass for
particular choices of soft terms.

Let us finally comment that our results do not directly apply to the case of Split SUSY \cite{splitsusy,Arvanitaki:2012ps}
in which one has $M,\mu,\ll  m$, since then the effect of light  gauginos and Higgssinos should be included in the
running below $M_{SS}$. In that case however it has been shown (see e.g.\cite{Cabrera:2011bi,Giudice:2011cg,Arbey:2011ab})
that split SUSY is only consistent with a 126 GeV Higgs for $M_{SS}\lesssim 100$ TeV and no intermediate scale scenario is possible.
Essentialy Split SUSY becomes a fine-tuned version of the standard MSSM.
One relevant issue is also that in Split SUSY, due to the smallness of gaugino masses,  in running down from the unification scale the scalar quarks of the third generation 
may easily become tachyonic, which restricts a lot the structure of the possible underlying SUSY breaking terms 
\cite{Arvanitaki:2012ps}.


\section{Discussion}

In this paper we have argued that the evidence found at LHC for a Higgs-like particle around $m_H\simeq 126$ GeV supports  the idea of
an underlying Supersymmetry being present at some ({\it not necesarily low}) mass scale. Even if the SUSY breaking scale is high, 
SUSY identities for dimension four operators remain true to leading order. In particular, the quartic
Higgs self-coupling $\lambda$ is related to the EW gauge couplings at the SUSY breaking scale, yielding constraints on the 
Higgs mass. The presence of SUSY restores the stability of the SM Higgs potential which otherwise becomes unstable at
high scales.

It is remarkable that the simple assumption of Higgs mass
parameter unification $m_{H_u}=m_{H_d}$ at the unification scale and minimal fine-tuning directly predict a Higgs mass in the range 
$m_H=126\pm 3$ GeV, consistent with LHC results,  for a SUSY scale $\gtrsim 10^ {10}$ GeV. For smaller values of $M_{SS}$  
the Higgs mass tends  to the value of a standard fine-tuned MSSM scenario with $m_H\lesssim 130$ GeV. 
Both situations with (relatively) low and High scale SUSY are consistent with the Higgs data (see e.g. fig.\ref{plotmas2}). 
Since in the context of the SM {\it any value} from e.g 100 GeV to 1 TeV would have been possible, one may interpret
this result as indirect evidence for an underlying SUSY. 

It has been argued that the fact that $\lambda \simeq 0$ near the Planck scale and that the  SM Higgs potential 
seems to be close to metastability could have some deep meening \cite{piraos} .
In our setting with SUSY at a large scale the quartic coupling $\lambda$,  is always 
positive definite and no such instability arises. The smallness of $\lambda$ is due to the fact that the EW couplings $g_1,g_2$ are small
and that the boundary condition $m_{H_u}=m_{H_d}$ at the unification scale keeps tan$\beta$ close to one for $M_{SS}\gtrsim 10^ {10}$ GeV.
As discussed in the previous section, such a situation with an intermediate scale SUSY breaking  and gauge coupling unification
may be naturally embedded in string theory compactifications like those resulting from F-theory $SU(5)$ unification.  The embedding into string theory is
also suggested  in order to understand the required fine-tuning in terms of the string landscape of compactifications.

 LHC at  13 TeV will be able to test the low SUSY breaking regime for squark and gluino masses of the order of a few TeV. If no direct trace of SUSY or any
other alternative new physics  is found at LHC, 
the case for a fine-tuning/landscape approach to the hierarchy problem will become stronger.
Still, as we have argued, heavy SUSY may be required for the stability of the Higgs potential and we have shown that the 
value $m_H\simeq 126$ GeV is  generic for $M_{SS}\gtrsim 10^ {10}$ GeV (see e.g. figures \ref{plotguay} and \ref{plotmass6}),
hinting to   a heavy SUSY scale.

One apparent shortcoming of  High scale SUSY is that we lose  the posibility of using the lightest neutralino as a dark matter candidate.
In this context 
the case of an intermediate scale  SUSY breaking
$M_{SS}\simeq 10^ {10}$ GeV is particularly interesting. Indeed, as recently discussed in \cite{imrv} (see also \cite{Chatzistavrakidis:2012bb,Giudice:2012zp}),
such scale may be compatible with an axion with decay constant $F_a\simeq 10^ {12}$ GeV, appropriate to provide for the required dark matter in
the universe. 
Furthermore, gauge coupling unification 
may elegantly be accomodated due to the presence of small threshold corrections as discussed in \cite{imrv}. 

Although a large scale for SUSY makes it difficult to test this idea directly at accelerators, indirect evidence could
be obtained. 
Improved precission on the measured values of both the top quark and the Higgs masses (e.g. at a linear collider) can make the 
constraints on specific High SUSY breaking models and the Higgs mass predictions more precise, along the lines discussed in the previous section.
Going beyond the next to leading order in the Higgs mass computation would also be required, see \cite{espinosa}.
If those measurements were precise enough, specific choices of soft terms and SUSY breaking scenarios could be ruled out or in.
Additional evidence in favour of an intermediate scale SUSY secenario could come from dark matter axion detection in microwave cavity search experiments like ADMX
\cite{Asztalos:2009yp}. Furthermore, since in models with large SUSY breaking scale the unification scale typically decreases, 
proton decay rates could also be at the border of detectability \cite{imrv}.  Finally, if any deviation from the SM expectations 
is found at low energies (like e.g. an enhanced Higgs rate to $\gamma \gamma$) the idea of a large SUSY scale  with a
fine-tuned Higgs would be immediately ruled out.

 \vspace{1.5cm}

\bigskip

\centerline{\bf \large Acknowledgments}

\bigskip

We thank  P.G.~C\'amara, A. Casas, J.R. Espinosa and F. Marchesano for useful discussions. 
This work has been partially supported by the grants FPA 2009-09017, FPA 2009-07908, Consolider-CPAN (CSD2007-00042) from the MICINN, HEPHACOS-S2009
/ESP1473 from the C.A. de Madrid and the contract ``UNILHC" PITN-GA-2009-237920 of the European Commission. We also thank the 
spanish MINECO {\it Centro de excelencia Severo Ochoa Program} under grant SEV-2012-0249.

\newpage

\appendix

\section{Renormalization group equations\label{RGE}}

Here we first  present the renormalization group equations at two loops for the SM couplings (the three gauge couplings, the top Yukawa and the Higgs quartic coupling).

\beqa
\frac{dg_1}{dt}&=&\frac{1}{(4 \pi )^2}\frac{41}{6}g_1^3+\frac{g_1^3}{(4 \pi )^4} \left(\frac{199}{18} g_1^2+\frac{27}{6} g_2^2+\frac{44}{3} g_3^2-\frac{17}{6} h_t^2\right)\\
\label{g1}
\frac{dg_2}{dt}&=&-\frac{1}{(4 \pi )^2}\frac{19 }{6 }g_2^3+\frac{g_2^3}{(4 \pi )^4} \left(\frac{9}{6} g_1^2+\frac{35 }{6}g_2^2+12 g_3^2-\frac{3 }{2}h_t^2\right)\\
\label{g2}
\frac{dg_3}{dt}&=&-\frac{1}{(4 \pi )^2}7g_3^3+\frac{g_3^3}{(4 \pi )^4} \left(\frac{11  }{6}g_1^2+\frac{9 }{2}g_2^2-26 g_3^2-2 h_t^2\right)
\label{g3}
\eeqa
\beqa
\frac{dh_t}{dt}&=&\frac{1}{(4 \pi )^2}h_t \left(\frac{9 h_t^2}{2}-\frac{17 g_1^2}{12}-\frac{9 g_2^2}{4}-8 g_3^2\right)+\frac{1}{(4 \pi )^4}h_t \left(-12 h_t^4+\frac{6 \lambda ^2}{4}-\right.\nonumber\\
& &\left.-\frac{12}{2} \lambda  h_t^2+\frac{131}{16} g_1^2 h_t^2+\frac{225}{16} g_2^2 h_t^2+36 g_3^2 h_t^2+\frac{1187}{216}  g_1^4-\frac{23 g_2^4}{4}-\right.\nonumber\\
& & \left.-108 g_3^4-\frac{3}{4}g_1^2 g_2^2+9 g_2^2 g_3^2+\frac{19}{9}g_3^2 g_1^2\right)
\label{gt}
\eeqa
\beqa
\frac{d\lambda}{dt}&=&\frac{1}{(4 \pi )^2}\left(12 \lambda  h_t^2-9 \lambda  \left(\frac{g_1^2}{3}+g_2^2\right)-4 3 h_t^4+\frac{3}{4} g_1^4+\frac{3}{2} g_2^2 g_1^2+\frac{9}{4} g_2^4+12 \lambda ^2\right)+ \nonumber\\
& &+\frac{1}{(4 \pi )^4}2 \left(-\frac{312}{8} \lambda ^3+\frac{36}{4} \lambda ^2 \left(g_1^2+3 g_2^2\right)-\frac{1}{2} \lambda  \left(-\frac{629}{24}  g_1^4-\frac{39}{4}g_1^2 g_2^2+\right.\right. \nonumber\\
& &\left.+\frac{73 g_2^4}{8}\right)+\frac{305 g_2^6}{16}-\frac{289}{48}  g_1^2 g_2^4-\frac{559}{48} g_1^4 g_2^2-\frac{379}{48} g_1^6-32 g_3^2 h_t^4-\nonumber\\
& &\left.-\frac{8}{3} g_1^2 h_t^4-\frac{9}{4} g_2^4 h_t^2+\frac{1}{2} \lambda  h_t^2 \left(\frac{85}{6}g_1^2+\frac{45 g_2^2}{2}+80 g_3^2\right)+\right.\nonumber\\
& &\left.+g_1^2 h_t^2 \left(-\frac{19}{4} g_1^2+\frac{21 g_2^2}{2}\right)-\frac{144}{4} \lambda ^2 h_t^2-\frac{3}{2} \lambda  h_t^4+30 h_t^6\right)
\label{l}
\eeqa

And finally the RGE (at 2 loops for gauge couplings, leading order in $h_t$) for the SUSY case:

\beqa
\frac{dg_1}{dt}&=&\frac{11 g_1^3}{ (4 \pi )^2}+\frac{g_1^3}{(4 \pi )^4} \left(\frac{199}{9}  g_1^2+9 g_2^2+\frac{88}{3} g_3^2-\frac{26}{3}  h_t^2\right)\label{g1susy}\\
\frac{dg_2}{dt}&=&\frac{g_2^3}{(4 \pi )^2}+\frac{g_2^3}{(4 \pi )^4} \left(3 g_1^2+25 g_2^2+24 g_3^2-6 h_t^2\right)\label{g2susy}\\
\frac{dg_3}{dt}&=&-\frac{3 g_3^3}{(4 \pi )^2}+\frac{g_3^3}{(4 \pi )^4} \left(\frac{11}{3}  g_1^2+9 g_2^2+14 g_3^2-4 h_t^2\right)\label{g3susy}\\
\frac{dh_t}{dt}&=&\frac{h_t}{(4 \pi )^2} \left(6 h_t^2-\frac{13 g_1^2}{9}-3 g_2^2-\frac{16 g_3^2}{3}\right)
\label{gtsusy}
\eeqa

\section{RGE solutions for the soft terms}

Here we display all the functions that appear in the solution of the RGE for the Higgs mass parameters $m_{Hu}$ and $m_{Hd}$ (see ref.\cite{ilm}).

First we define the functions
\beq
E(t)\ =\ (1+\beta_3t)^{16/(3b_3)}(1+\beta_2t)^{3/(b_2)}(1+\beta_1t)^{13/(9b_1)}
\ \ ,\ \ 
F(t)=\int_0^t E(t')dt'
\eeq
with $\beta_i=\alpha_i(0)b_i/(4\pi)$
and $t=2\log(M_c/M_{SS})$. 
The beta-functions coefficients for the SUSY case are $(b_1,b_2,b_3)=(11,1,-3)$ and we define $\alpha_0=\alpha(0)=\alpha_i(0)=g_i^2(0)/(4\pi^2)$ for $i=2,3$,
 $\alpha_1(0)=(3/5)\alpha(0)=g_1^2(0)/(4\pi^2)$ where $\alpha_0$ is the unified coupling at $M_c$.
In our case the couplings do not strictly unify, only up to $5\%$ corrections.
In the numerical computations we take the average value of the 
three couplings at $M_c$, which is enough for our purposes.

We then define the functions in eqs.(\ref{ilm2})
\beqa
q(t)^2  & =&  \frac {1}{(1+6Y_0F(t))^{1/2}}(1+\beta_2t)^{3/b_2}
(1+\beta_1t)^{1/b_1} \ ;\ 
h(t) = \frac {1}{2}(3/D(t)-1)  \nonumber \\ 
k(t)\ & = & \ \frac {3Y_0F(t)}{D(t)^2} \ ;\
f(t)\ =\ -\frac {6Y_0H_3(t)}{D(t)^2} \ ;\ D(t)\ =\ (1+6Y_0F(t)) \\ \nonumber
e(t)\ & =& \ \frac {3}{2} \left( \frac {(G_1(t)+Y_0G_2(t))}{D(t)} 
\ +\ \frac {(H_2(t)+6Y_0H_4(t))^2}{3D(t)^2} \ +\ H_8\right) \nonumber 
\eeqa
where $Y_0=Y_t(0)$ and $Y_t=h_t^2/(4\pi)^2$. 
The functions $g,H_2,H_3,H_4,G_1,G_2$ and $H_8$ are   independent of the top Yukawa  coupling, only depend on the gauge coupling constants and are given by

\beqa
g(t) \  &=&\ \frac {3}{2} \frac {\alpha_2(0)}{4\pi} f_2(t) \  + \ \frac {1}{2} \frac {\alpha_1(0)}{4\pi}
f_1(t) \nonumber \\
H_2(t) \ & =& \  \frac {\alpha_0}{4\pi}\left(\frac {16}{3}h_3(t)\ +\ 3h_2(t)\ +\ \frac {13}{15}h_1(t) \right) \nonumber \\
H_3(t) \  &=& \ tE(t)\ -\ F(t) \nonumber \\
H_4(t)\ &=& \ F(t)H_2(t) \ -\ H_3(t) \nonumber \\
H_5(t) \  & = & \ \frac {\alpha_0}{4\pi} \left(-\frac{16}{3}f_3(t) \ +\ 6 f_2(t) \ -\ \frac {22}{15} f_1(t) 
\right) \nonumber \\
H_6(t) \ & =& \ \int_0^t\ H_2(t')^2 E(t')dt' \nonumber \\
H_8(t) \ & = & \ \frac {\alpha_0}{4\pi }\ \left( -\frac {8}{3}f_3(t)\ +\ f_2(t) \ -\ \frac {1}{3} f_1(t) \right) \nonumber
\eeqa
\beqa
G_1(t) \ & = &\ F_2(t) \ -\ \frac {1}{3} H_2(t)^2 \nonumber \\
G_2(t) \ & = & \ 6F_3(t) \ -\ F_4(t) \ -\ 4H_2(t)H_4(t) \ +\ 2F(t)H_2(t)^2 \ 
\ -\ 2H_6(t) \nonumber \\
F_2(t) \ & = & \ \frac {\alpha_0}{4\pi } \ \left( \frac {8}{3}f_3(t) \ +\ \frac {8}{15}f_1(t) \right)
\nonumber \\
F_3(t) \ & = &\ F(t)F_2(t)\ -\ \int_0^t E(t')F_2(t') dt' \nonumber \\
F_4(t) \ & = & \ \int_0^t E(t')H_5(t') dt' 
  \eeqa
  where $f_i(t)$ and $h_i(t)$ are defined by
  \beq 
  f_i(t) \ =\ \frac {1}{\beta_i}(1\ -\ \frac {1}{(1+\beta_it)^2}) 
  \ ;\ 
  h_i(t) \ =\ \frac {t}{(1+\beta_it)} \ .
  \eeq
   The low energy of the top mass may be obtained from 
   the solutions of the one-loop renormalization group equations,
   devided into two pieces, SUSY and non-SUSY,  i.e. (here $Y_t=h_t^2/(16\pi^2)$)
\beq
Y_t(m_t) \ =\ sin^2\beta Y_t(M_{SS})  \frac  {E'(t_{EW})}{(1+(9/2) sin^2\beta Y_t(M_{SS})F'(t_{EW}))}
\eeq
where
\beq
Y_t(M_{SS}) \ =\   Y_t(M_c)  \frac  {E(t_{SS})}{(1+6  Y_t(M_c)F(t_{SS}))} 
\label{topyukis}
  \eeq
The functions $E,F$ are as defined above, with $t_{SS}=2log(M_c/M_{SS})$
and $t_{EW}=2log(M_{SS}/M_{EW})$, while the functions $E',F'$ are analogous to $E,F$ but replacing the $b_i$ and anomalous dimensions
by the non-SUSY ones, i.e. 
\beq
E'(t)\ =\ (1+\beta'_3t)^{8/(b^{NS}_3)}(1+\beta'_2t)^{9/(4b^{NS}_2)}(1+\beta'_1t)^{17/(12b^{NS}_1)}
\ \ ,\ \ 
F'(t)=\int_0^t E'(t')dt'
\eeq
with $\beta'_i=\alpha_i(M_{SS})b^{NS}_i/(4\pi)$, $b_i^{NS}=(41/6,-19/6,-7)$ and $t=t_{EW}$. 
For the anomalous dimensions we have made the change in the definition of $E(t)$ 
$(13/9,3,16/3)$ $\rightarrow(17/12,9/4,8)$.
And we take the value of $h_t(m_t)$ computed in eq.(\ref{gtm}) taking into account the threshold corrections at electroweak scale. For this particular computation we take actually as electroweak scale the top mass, so $t_{EW}=2log(M_{SS}/m_t)$.\\

Finally, in order to compute the value of the stop mixing parameter $X_t$ we need the following equations for the running of the soft parameters:
\beqa
A_t(t)&=&\frac{A}{D(t)}+M(H_2(t)-\frac{6Y_0H_3(t)}{D(t)})\nonumber \\
\mu(t)&=&\mu_0 q(t)\nonumber \\
m_4^2(t)&=&M^2(-3\frac{\alpha_2(0)}{4\pi}f_2(t)+\frac{\alpha_1(0)}{4\pi}f_1(t))\nonumber \\
m_5^2(t)&=&-\frac{1}{3}m^2+M^2(-\frac{8}{3}\frac{\alpha_3(0)}{4\pi}f_3(t)+\frac{\alpha_2(0)}{4\pi}f_2(t)-\frac{5}{9}\frac{\alpha_1(0)}{4\pi}f_1(t))\nonumber \\
m_D^2(t)&=&m^2+2M^2(\frac{4}{2}\frac{\alpha_3(0)}{4\pi}f_3(t)+\frac{3}{4}\frac{\alpha_2(0)}{4\pi}f_2(t)+\frac{1}{36}\frac{\alpha_1(0)}{4\pi}f_1(t))\nonumber \\
m_U^2(t)&=&\frac{2}{3}m_{Hu}^2(t)-\frac{2}{3}\mu^2(t)-m_5^2(t)\nonumber \\
m_Q^2(t)&=&\frac{1}{2}m_D^2(t)-\frac{1}{2}m_4^2(t)+\frac{1}{2}m_U^2(t)
\eeqa 
   
 \section{Threshold corrections at the EW scale}

The functions appearing in the computation of the threshold corrections to the Higgs self-coupling  at the weak scale are given by 
 \cite{Sirlin:1985ux}:
\begin{eqnarray}
F_1&=&12 \log\left[\frac{Q}{m_h}\right]+\frac{3 \log[\xi ]}{2}-\frac{1}{2} Z\left[\frac{1}{\xi }\right]-Z\left[\frac{c_W^2}{\xi }\right]-\log\left[c_W^2\right]+\frac{9}{2} \left(\frac{25}{9}-\frac{\pi }{\sqrt{3}}\right)
\label{f1}
\end{eqnarray}
\begin{eqnarray}
F_0&=&-12 \log\left[\frac{Q}{M_Z}\right] \left(1+2 c_W^2-\frac{2 m_t^2}{M_Z^2}\right)+\frac{3 c_W^2 \xi  \log\left[\frac{\xi }{c_W^2}\right]}{\xi -c_W^2}+2 Z\left[\frac{1}{\xi }\right]+4 c_W^2 Z\left[\frac{c_W^2}{\xi }\right]+\nonumber\\
& &+\frac{3 c_W^2 \log\left[c_W^2\right]}{s_W^2}+12 c_W^2 \log\left[c_W^2\right]-\frac{15}{2} \left(1+2 c_W^2\right)-\nonumber\\
& &-\frac{3 m_t^2 \left(2 Z\left[\frac{m_t^2}{M_Z^2 \xi }\right]-5+4 \log\left[\frac{m_t^2}{M_Z^2}\right]\right)}{M_Z^2}
\label{f0}
\end{eqnarray}
\begin{eqnarray}
F_3&=&12 \log\left[\frac{Q}{M_Z}\right] \left(1+2 c_W^4-\frac{4 m_t^4}{M_Z^4}\right)-6 Z\left[\frac{1}{\xi }\right]-12 c_W^4 Z\left[\frac{c_W^2}{\xi }\right]-12 c_W^4 \log\left[c_W^2\right]+\nonumber\\
& &+8 \left(1+2 c_W^4\right)+\frac{24 m_t^4}{M_Z^4} \left(Z\left[\frac{m_t^2}{M_Z^2 \xi }\right]-2+\log\left[\frac{m_t^2}{M_Z^2}\right]\right)
\label{f3}
\end{eqnarray}
where $\xi=m_h^2/M_Z^2$, $c_W=\cos \theta_W$, $s_W=\sin \theta_W$ and
\begin{equation}
Z(z)=\left\{\begin{array}{lcc}2 \zeta \arctan\left[\frac{1}{\zeta}\right]& for& z>1/4\\
\zeta \log[\frac{1+\zeta}{1-\zeta}]&for & z<1/4\end{array}\right.
\end{equation}
where $\zeta=\sqrt{Abs[1-4 z]}$.
In the computation we have taken the central experimental values for $M_Z$, $m_t$ and $s_W$ given by eqs.(\ref{input1},\ref{input2}) and the tree level value for the Higgs mass, i.e. $m_h^2=2\lambda v^2$ with $v=174.1$.

\end{document}